\documentclass[aps,prb,bibnotes,twocolumn,preprintnumbers,amsmath,amssymb,superscriptaddress,floatfix]{revtex4-2}

\usepackage[T1]{fontenc}
\usepackage{graphicx}
\usepackage{dcolumn}
\usepackage{bm}
\usepackage{amsthm}
\usepackage{amsmath}
\usepackage{amssymb}
\usepackage{hyperref}
\usepackage{url}
\usepackage{xcolor}
\usepackage{comment}

\def\afflux{Department of Physics and Materials Science, University of Luxembourg, 162A~Avenue de la Faiencerie, L-1511 Luxembourg, Grand Duchy of Luxembourg}
\def\affias{Institute for Advanced Studies, University of Luxembourg, Campus Belval, L-4365 Esch-sur-Alzette, Grand Duchy of Luxembourg}
\def\affghent{Department of Solid State Sciences, Ghent University, Krijgslaan 285/S1, 9000 Ghent, Belgium}
\def\affmpg{Max Planck Institute for the Structure and Dynamics of Matter, Luruper Chaussee~149, D-22761 Hamburg, Germany}
\def\affcfel{Center for Free-Electron Laser Science, Notkestra\ss e~85, D-22607 Hamburg, Germany}

\begin{document}

\title{Reduced vortex descriptors linking polycrystallinity in magnetic nanoparticles with polarized magnetic small-angle neutron scattering}

\author{Michael P.\ Adams}\email{michael.adams@mpsd.mpg.de}
\affiliation{\affmpg}
\affiliation{\affcfel}
\author{Jonathan Leliaert}
\affiliation{\affghent}
\author{Andreas Michels}
\affiliation{\afflux}
\author{Elizabeth M.\ Jefremovas}\email{elizabeth.jefremovas@uni.lu}
\affiliation{\afflux}
\affiliation{\affias}

\date{\today}

\begin{abstract}
Analytical vortex models reduce polarized magnetic small-angle neutron scattering (SANS) from nanoparticle ensembles to a small set of texture descriptors. In this work, we apply this reduction to micromagnetic simulations of polycrystalline iron oxide nanoflowers at a fixed particle size and examine how a controlled parametrization of multigrain disorder is reflected in the remanent descriptors. The particles are represented by explicit Voronoi microstructures, and intraparticle disorder is varied through the intergrain exchange coupling and anisotropy-axis coherence. Fitting each remanent magnetization state to a hyperbolic vortex model reveals a predominantly two-channel organization: the intergrain exchange coupling is associated mainly with the radial vortex profile, whereas the anisotropy-axis coherence is associated mainly with the orientational moment of the vortex-axis distribution. The normalized spin-flip SANS cross sections are accurately represented by independent fits of the analytical linear-vortex SANS expression obtained from the first-order expansion of the hyperbolic profile. The fitted orientational descriptor agrees closely with its independent real-space estimate, whereas the corresponding radial descriptors exhibit a strong global nonlinear relation. This separation identifies which information from the micromagnetic vortex textures is robustly retained by the reduced analytical representation.
\end{abstract}

\maketitle


\section{Introduction}

Magnetic nanoparticles (MNPs) are often treated within the macrospin approximation as single-domain, uniformly magnetized entities~\cite{nowak2005spin}. This description is often adequate for sufficiently small particles (typically tens of nanometers~\cite{jefremovas2021nanoflowers,mekseriwattana2025,borchers2025magnetic}), but becomes insufficient in the intermediate-size regime, where microstructural defects and associated magnetic disorder produce complex, nonuniform magnetization textures~\cite{lu2007,gazeau2012No1,weiyang2015,lappas2019,lakbenderdisch2021,prajapati2024,eliss2026,perspectives}.

Multicore iron oxide MNPs (also known as nanoflowers, NFs) constitute a model example of a defect-rich nanoarchitecture. Their excellent macroscopic heat dissipation, key in applications such as magnetic hyperthermia or nanocatalysis~\cite{gavilan2025magnetic,gazeau2012No1,zanette2025carbon}, is argued to be a direct consequence of their internal magnetic disorder~\cite{bender2018dipolar,bender2018relating,eliss2026,jefremovas2026heating}. Their complex, nonuniform intraparticle magnetization states result from the interplay among the crystallographic cores, their anisotropy-axis disorder and pinning landscape, and the exchange and dipolar fields. Elucidating the resulting distribution of magnetic moments is therefore essential for connecting their synthesis-controlled structure with their macroscopic magnetic performance.

Polarized small-angle neutron scattering (SANS) is well positioned to address this challenge. This reciprocal-space technique probes statistically-averaged nanoscale magnetization correlations in bulk amounts of material~\cite{rmp2019,michelsbook}. It has been used to resolve magnetic correlations in MNP assemblies, including NFs, for which dipolar-coupled moment and supraferromagnetic correlations have been reported~\cite{bender2018dipolar,benderapl2019,honecker2020}. More recently, nonuniform vortexlike correlations in maghemite multicore aggregates have been studied, illustrating the potential and sensitivity of polarized SANS to elucidate magnetic correlations associated with structural features~\cite{rai2026}.

This sensitivity also exposes a central ambiguity of the inverse problem: different real-space magnetization textures or microscopic disorder parameters may produce similar SANS anisotropies. A useful strategy is therefore to combine micromagnetic simulations with analytical vortex models and to reduce the SANS response of vortex-state particles to a small number of collective variables.

Previous works established analytical spin-flip SANS expressions for linear vortex magnetization fields and obtained the angular anisotropy landscape generated by the radial vortex profile and the vortex-axis distribution~\cite{adamsprb2024no2,adams2026angular}. A related minimal vortex model showed that the radial profile parameter \(\nu\) is itself a field-dependent collective coordinate along the hysteresis loop of a spherical, structurally uniform nanoparticle~\cite{adamsprb2026minimal}. In those earlier examples, simple axis-distribution models were sufficient to describe the features under consideration. Polycrystalline multicore nanoarchitectures, such as the NFs studied here, introduce two further distinctions within the linear-vortex description: (i)~the polar axis distribution enters the spin-flip SANS cross section through the reduced orientational moment \(\Lambda\); and (ii)~reconstructing the full vortex-axis distribution imposes an additional model assumption. The linear-vortex representation is particularly useful for the reciprocal-space reduction because its polynomial spatial form permits a closed analytical Fourier transform and orientational ensemble average. Thus, the vortex-SANS problem is examined through two complementary reductions. The micromagnetic textures are fitted in real space to obtain radial and orientational vortex descriptors, whereas the numerical SANS cross sections are fitted independently using the analytical linear-vortex expression. Comparing the resulting descriptors tests which information is represented consistently in the real-space and reciprocal-space descriptions.


In this work, we use polycrystalline iron oxide NFs as a forward model for this descriptor-mapping problem. Synthesis studies have demonstrated chemical control over mesocrystal properties, including their size and degree of crystallinity~\cite{gavilan2017,jefremovas2021nanoflowers,gavilan2021, gavilan2025magnetic,simeonidis2024toward}. Here, we represent the resulting magnetic microstructure through a controlled interpolation between random and coherently biased grain-scale anisotropy landscapes, together with a variable intergrain exchange coupling. These parameters define a controlled micromagnetic model rather than direct experimental measures of crystallinity or grain-boundary coupling. Moreover, it has recently been shown that the degree of intraparticle disorder critically affects the macroscopic properties and can promote vortexlike remanent states~\cite{eliss2026,jefremovas2026heating}, underscoring the importance of microstructural control as a route to tailoring particle functionality. We herein harness such a polycrystalline nanoarchitecture as a model system to evaluate how the controlled multigrain parametrization is reflected in the collective vortex variables that enter the analytical vortex-SANS response.

Our study is therefore deliberately restricted to a controlled remanent configuration: the particles are saturated along \(+\mathbf{e}_z\), relaxed, the field is set to zero, and the system is relaxed again (using MuMax3~\cite{mumax3new}). At this single field point, we vary two model parameters: (i)~the intergrain exchange scale \(k\); and (ii)~the fraction \(f\) of grains whose anisotropy axes are aligned with \(\mathbf{e}_z\). The resulting numerical and reduced-analysis routes can be summarized as follows:x
\begin{widetext}
\[
\begin{array}{@{}c@{\quad}c@{\quad}c@{\quad}c@{\quad}c@{}}
\overset{
\substack{
\mathrm{micromagnetic}\\
\mathrm{model\ point}
}
}{
\bigl(f,k\mid R_{\mathrm{nom}}\bigr)
}
&
\xrightarrow[
\mu_0H_0:1.2\,\mathrm{T}\rightarrow 0\,\mathrm{T}
]{\mathrm{MuMax3\;relax}}
&
\overset{
\substack{
\mathrm{ensemble\ of\ discrete}\\
\mathrm{magnetization\ textures}
}
}{
\left\{
\mathbf{M}_j(\mathbf{r}_i;H_0=0)
\right\}_{j=1}^{N}
}
&
\xrightarrow{\mathrm{NuMagSANS}}
&
\overset{
\substack{
\mathrm{ensemble\mbox{-}averaged}\\
\mathrm{spin\mbox{-}flip\ SANS}\\
\mathrm{cross\ section}
}
}{
\left\langle
\frac{d\Sigma_{\mathrm{sf}}}{d\Omega}
\right\rangle_{\mathrm{num}}
}
\\[0.8em]
&&
\underset{
\substack{
\mathrm{object\mbox{-}wise}\\
\mathrm{hyperbolic\mbox{-}vortex\ fits}
}
}{\Big\downarrow}
&&
\underset{
\substack{
\mathrm{linear\mbox{-}vortex}\\
\mathrm{SANS\ fit}
}
}{\Big\downarrow}
\\[1.2em]
&&
\begin{gathered}
\textstyle \left\{\mathbf{M}_{\mathrm{hv},j}(\mathbf{r})\right\}_{j=1}^{N}\\
\textstyle \{\nu_j,\chi_j,\alpha_j,\beta_j;R_{\max,j}\}_{j=1}^{N}
\end{gathered}
&
\underset{
\substack{
\mathrm{analytical\ Fourier\ transform}\\
\mathrm{and\ ensemble\ average}
}
}{
\overset{
\mathrm{first\mbox{-}order\ expansion}
}{\Longrightarrow}
}
&
\begin{gathered}
\textstyle
\left\langle\frac{d\Sigma_{\mathrm{sf}}}{d\Omega}\right\rangle_{\mathrm{lv}}\\
\textstyle
(\nu_{\mathrm{rms,fit}},\Lambda_{\mathrm{fit}};R_{\mathrm{fit}})
\end{gathered}
\\[0.8em]
&&
\underset{\mathrm{ensemble\mbox{-}averaged\ descriptors}}{\Big\downarrow}
&&
\underset{\mathrm{fit\ parameters}}{\Big\downarrow}
\\[1.2em]
&&
\underbrace{
\left(
\nu_{\mathrm{rms},N},
\Lambda_N;
\langle R_{\max}\rangle_{N}
\right)
}_{\mathrm{real\mbox{-}space}}
&
\overset{
\substack{
\Lambda:\ \mathrm{near\ equality}\\
\nu/R:\ \mathrm{nonlinear\ relation}
}
}{\longleftrightarrow}
&
\underbrace{
\left(
\nu_{\mathrm{rms,fit}},
\Lambda_{\mathrm{fit}};
R_{\mathrm{fit}}
\right)
}_{\mathrm{SANS\ fit}} .
\end{array}
\]
\end{widetext}
The two vertical branches represent independent reductions of the same micromagnetic ensemble. The real-space branch fits each magnetization texture object-wise with the hyperbolic vortex model and forms the corresponding ensemble descriptors. The reciprocal-space branch first evaluates the spin-flip SANS cross section directly from the full magnetization fields using the open-source software package NuMagSANS~\cite{adamsjac2026} and then fits the resulting numerical pattern with the linear-vortex SANS expression. The double-lined horizontal arrow denotes the analytical model relation: the hyperbolic profile is expanded to first order, Fourier transformed, and ensemble averaged to obtain the linear-vortex SANS expression. It does not denote a direct parameter transfer between the two fits. Their comparison instead reveals which reduced vortex properties are represented consistently in both descriptions. The present work does not attempt a full field-dependent parameter-space screening. Instead, it examines a controlled multigrain micromagnetic model at remanence within the linear-vortex framework.

Section~\ref{methods} introduces the micromagnetic model, spin-flip SANS cross section, and analytical vortex reduction; Sec.~\ref{results} presents the resulting collective variables and 2D SANS patterns; and Sec.~\ref{conclusion} summarizes the findings. Further details on the simulation and analysis are given in the Supplemental Material~\cite{smelisim2026}.


\section{Methods}
\label{methods}

\subsection{Micromagnetic model}

Numerical micromagnetic simulations were performed using MuMax3~\cite{mumax3new}. The standard energy contributions consisting of isotropic exchange, (uniaxial) magnetocrystalline anisotropy, dipolar interaction, and Zeeman energy were taken into account. We consider NFs with a nominal diameter of $D_{\mathrm{nom}} = 2R_{\mathrm{nom}} = 100 \, \mathrm{nm}$ on a cubic mesh with $36 \times 36 \times 36$ cells and a cell-edge length of $a_{\mathrm{cell}} = 5 \, \mathrm{nm}$. This discretization is smaller than the exchange length $l_{\mathrm{ex}} = \sqrt{2A/(\mu_{0}M_{\mathrm{s}}^{2})} \simeq 10.0 \, \mathrm{nm}$ of maghemite. 

The polycrystalline microstructure was generated outside MuMax3 as a fixed three-dimensional (3D) Voronoi region map with a nominal mesocrystal size parameter \(g_{\mathrm{s}} = 15\,\mathrm{nm}\) (equivalently, a grain-length scale), and subsequently imported into the micromagnetic simulation. Surface grains were allowed to extend beyond the nominal spherical mask, while nonmagnetic voids inside the nominal sphere were excluded by construction. This procedure fixes the structural ensemble before the micromagnetic calculation and makes the realized grain geometry directly inspectable. The corresponding generation procedure, stored metadata, and validation checks are described in the Supplemental Material~\cite{smelisim2026}. The generated simulation data and the analysis workflow are archived as a Zenodo dataset~\cite{adams2026nanoflowerdataset}.

\begin{figure}[t!]
\centering
\includegraphics[width=1.0\columnwidth]{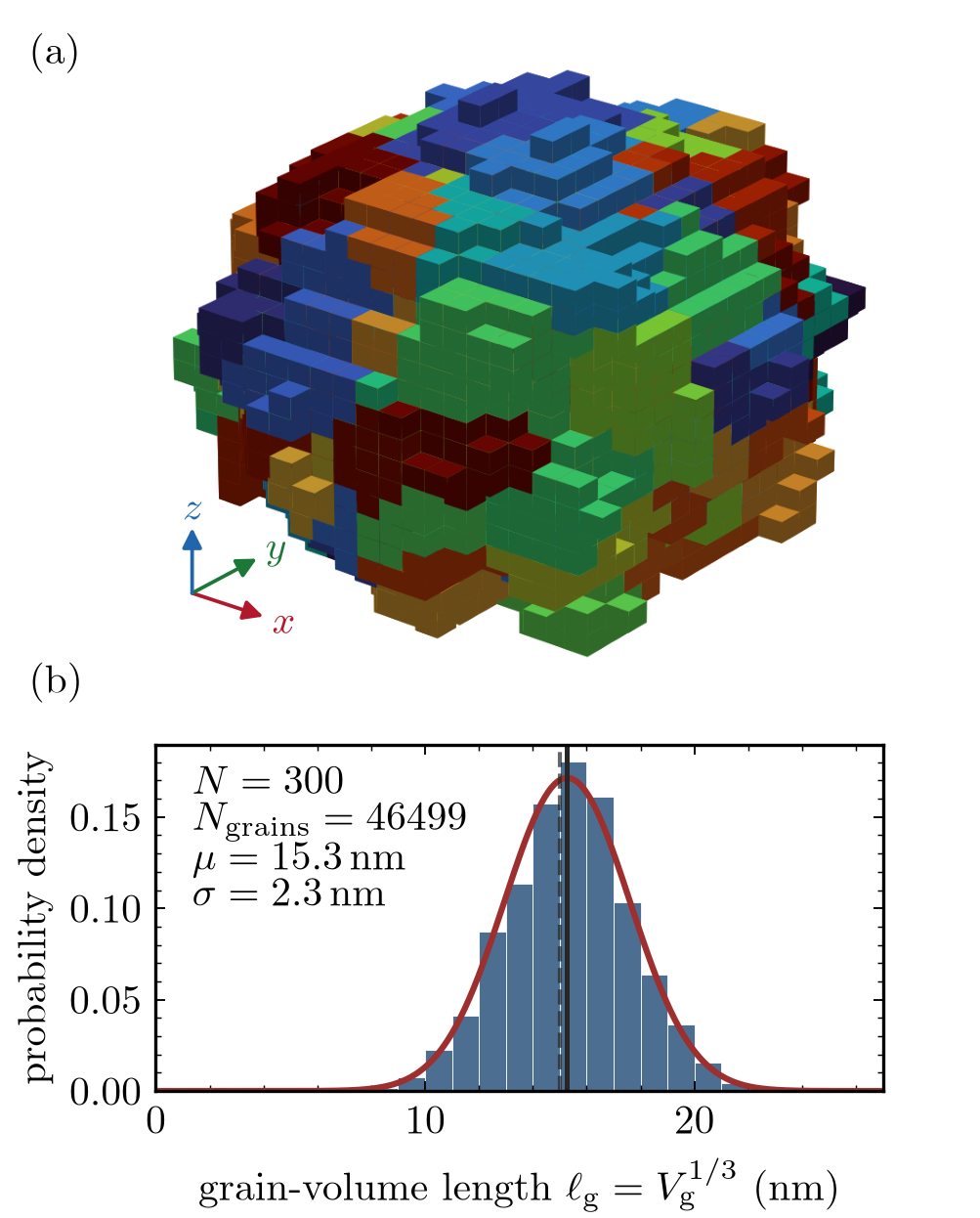}
\caption[Micromagnetic nanoflower model]{Example of the micromagnetic nanoflower (NF) model. (a)~Representative three-dimensional (3D) Voronoi realization used to define the polycrystalline grain structure of a single NF. The inset indicates the laboratory axes. (b)~Distribution of the volume-derived grain length \(\ell_{\mathrm{g}}=V_{\mathrm{g}}^{1/3}\), obtained by pooling the realized grain volumes over the simulated ensemble. The data were obtained for \(N=300\) NFs comprising a total of \(46\,499\) crystallites. The solid curve denotes a Gaussian fit with a mean of \(\mu=15.3\,\mathrm{nm}\) (solid vertical line) and a standard deviation of \(\sigma=2.3\,\mathrm{nm}\). The dashed line indicates the prescribed structural scale \(g_{\mathrm{s}}=15\,\mathrm{nm}\).}
\label{fig1}
\end{figure}

Figure~\ref{fig1}(a) shows one representative Voronoi realization used as a MuMax3 region map. The actual grain volumes were evaluated directly from the occupied simulation cells, and the volume-derived length \(\ell_{\mathrm{g}}=V_{\mathrm{g}}^{1/3}\) was calculated for each grain. The pooled distribution over the simulated NF ensemble is shown in Fig.~\ref{fig1}(b). It is approximately described by a Gaussian distribution with a mean of \(\mu=15.28\,\mathrm{nm}\) and a standard deviation of \(\sigma=2.33\,\mathrm{nm}\), consistent with the prescribed structural scale \(g_{\mathrm{s}}=15\,\mathrm{nm}\). The small upward shift reflects the uncut surface grains and the discrete mesh representation. The same metadata define the realized grain- and volume-weighted anisotropy fractions used below, so that both the structural and magnetic disorder are checked before entering the micromagnetic simulations.

To model different degrees of anisotropy-axis coherence, each grain $i$ was assigned a unit anisotropy axis $\mathbf{u}_{\mathrm{ani},i}$. A reference direction $\mathbf{u}_{\mathrm{ani},0}=\mathbf{e}_z$ was defined parallel to the saturating field direction. The dimensionless parameter $f$ denotes the fraction of grains whose anisotropy axes are aligned with this reference direction:
\begin{align}
f = \frac{N^{\parallel}_{\mathrm{G}}}{N_{\mathrm{G}}} ,
\end{align}
where \(N_{\mathrm{G}}\) is the number of grains in one NF and \(N^{\parallel}_{\mathrm{G}}\) is the number of those grains with \(\mathbf{u}_{\mathrm{ani},i}=\mathbf{u}_{\mathrm{ani},0}\). The total number of grains pooled over all \(N=300\) NFs is given only for the ensemble-level histogram in Fig.~\ref{fig1}(b). The remaining grains are assigned independent random unit vectors (sampled uniformly on the unit sphere). Consequently, \(f=1\) represents a fully coherent anisotropy-axis configuration, while \(f=0\) represents a random polycrystalline anisotropy-axis distribution. The parameter \(f\) is not intended as a direct experimental measure of crystallinity; it is a minimal control parameter for a uniaxial bias in the grain-scale anisotropy landscape. The anisotropy assignment is generated independently of the structural geometry by combining a fixed global anisotropy seed with the structural seed of each NF, and ensemble-averaged scattering observables are computed from the resulting set of realizations. The Supplemental Material~\cite{smelisim2026} gives the explicit sampling procedure, the realized grain- and volume-weighted values of \(f\), and the diagnostics used to exclude unintended correlations with the Voronoi geometry. 
 
Material parameters characteristic of maghemite were taken~\cite{eliss2026}: saturation magnetization $M_{\mathrm{s}} = 400 \, \mathrm{kA/m}$, exchange-stiffness constant $A = 10 \, \mathrm{pJ/m}$, and uniaxial anisotropy constant $K_{\mathrm{u}} = 10^4 \, \mathrm{J/m^{3}}$. The intragrain exchange stiffness was kept fixed at $A$, while the exchange coupling across the grain-boundary regions was rescaled by a dimensionless factor $k$ using the MuMax3 extension ``$\mathrm{ext\_ScaleExchange}$''. Thus, $k=1$ corresponds to fully exchange-coupled grains, whereas smaller values of $k$ represent weakened intergrain exchange coupling. 

For the selected material parameters, the homogeneous defect-free sphere reference estimate for the critical vortex-nucleation radius is $R_{\mathrm{nuc}} = \sqrt{15} l_{\mathrm{ex}} \left[1 - 6 \kappa^2\right]^{-\frac{1}{2}} \simeq 46.1\,\mathrm{nm}$, with a hardness parameter of $\kappa = \sqrt{K_{\mathrm{u}}/(\mu_0 M_{\mathrm{s}}^2)}\simeq0.223$~\cite{adamsprb2026minimal,adams2026minimalDataset}. The selected nominal NF radius of \(R_{\mathrm{nom}}=50\,\mathrm{nm}\) is close to this critical radius, with \(R_{\mathrm{nom}}/R_{\mathrm{nuc}}\simeq1.08\). This places the chosen size in a regime that is expected to be particularly sensitive to the competition between broad-core and flux-closure-dominated vortex textures. Because the simulated NFs are polycrystalline, this analytical value serves as a reference scale rather than an exact nucleation threshold.
For comparison, the \(R_{\mathrm{is}}=20\,\mathrm{nm}\) iron spheres used for the uniaxial micromagnetic validation of the linear-vortex SANS model in Ref.~\cite{adamsprb2024no2} correspond to \(R_{\mathrm{is}}/R_{\mathrm{nuc}}\simeq2.11\) for their respective material parameters. Those particles were therefore located substantially farther above the analytical nucleation scale than the present NFs, for which \(R_{\mathrm{nom}}/R_{\mathrm{nuc}}\simeq1.08\). The present simulations consequently probe a regime closer to the threshold, although the homogeneous defect-free sphere estimate $R_{\mathrm{nuc}}$ remains an approximate reference scale for the polycrystalline NFs.

Unless stated otherwise, the real-space vortex-parameter analysis, structural statistics in Fig.~\ref{fig1}(b), and the direct NuMagSANS~\cite{adamsjac2026} ensemble averages in Fig.~\ref{fig5}(a) use the simulated parameter grid \(f=\{0,0.1,\ldots,1.0\}\) and \(k=\{0.1,0.25,0.5,0.7,1.0\}\), with \(N=300\) independently generated NFs for each parameter pair. The best-fit reduced analytical patterns in Fig.~\ref{fig5}(b) are obtained independently from the normalized NuMagSANS cross sections, while Figs.~\ref{fig5}(c,d) compare the fitted parameters with the corresponding \(N=300\) real-space vortex descriptors. The ensemble is dilute in the sense that each NF is simulated independently; interparticle dipolar correlations are outside the scope of the present forward model. Here, \(k\) and \(f\) are micromagnetic model parameters; their relation to synthesis-controlled grain-boundary coupling or crystallinity is therefore model dependent.

All states analyzed in this work were generated using the same field protocol. The magnetization was initialized uniformly along the \(+\mathbf{e}_z\) direction. A saturating field of \(\mu_0 H_0 = 1.2\,\mathrm{T}\) was then applied along \(+\mathbf{e}_z\), and the system was relaxed. After this high-field relaxation, the external field was set to \(\mu_0 H_0 = 0\,\mathrm{T}\), followed by a second relaxation. The analyzed configuration is therefore a remanent state defined by this two-step field protocol, not a field-following state sampled along a complete hysteresis loop. This distinction is important because the radial vortex profile is in general field dependent~\cite{adamsprb2026minimal}. In the present work, we therefore examine how the selected polycrystalline parametrization is reflected in the remanent radial profile and vortex-axis distribution at this single field point. The energy functional, intergrain exchange scaling, and file-indexing protocol are summarized in the Supplemental Material~\cite{smelisim2026}.


\subsection{Spin-flip SANS cross section}

\begin{figure}[tb]
\centering
\includegraphics[width=1.0\columnwidth]{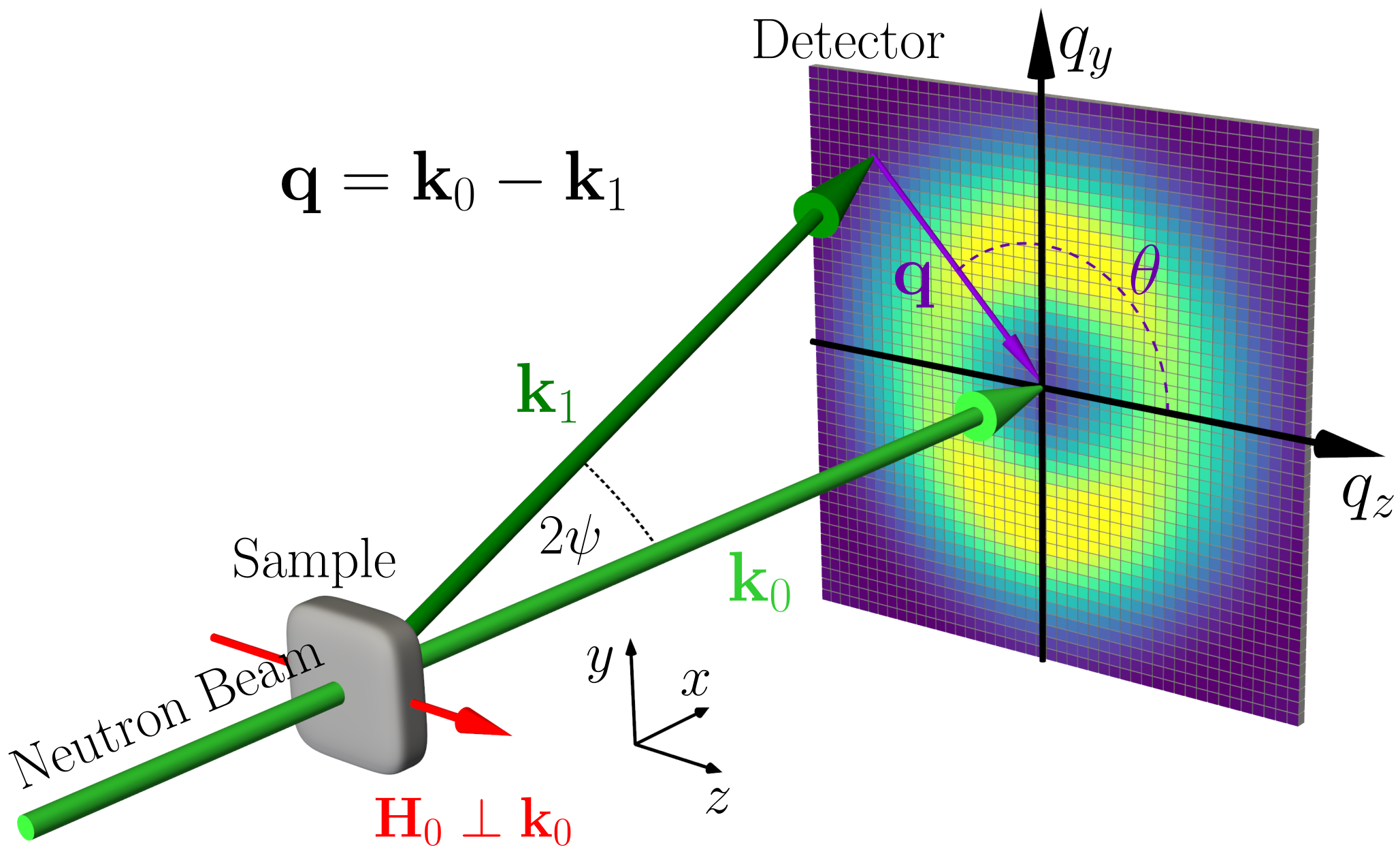}
\caption[Scattering geometry]{Scattering geometry used throughout this work. The applied magnetic (guide) field $\mathbf{H}_0 \parallel \mathbf{e}_z$ is perpendicular to the incident neutron wave vector $\mathbf{k}_{0} \parallel \mathbf{e}_x$, and the scattering vector $\mathbf{q}=\mathbf{k}_{0}-\mathbf{k}_{1}$ is detected in the $(q_y,q_z)$ detector plane. The azimuthal detector angle $\theta$ is measured with respect to the field direction.}
\label{fig2}
\end{figure}

The results for the spin-flip SANS cross section are calculated for the scattering geometry where the applied magnetic (guide) field $\mathbf{H}_0 \parallel \mathbf{e}_z$ is perpendicular to the incident neutron beam ($\mathbf{k}_0 \parallel \mathbf{e}_x$), as sketched in Fig.~\ref{fig2}. This configuration is commonly employed in uniaxial polarization analysis experiments. The corresponding elastic differential spin-flip SANS cross section $d\Sigma_{\mathrm{sf}} / d\Omega$ reads~\cite{michelsbook}:
\begin{equation}
\label{eq:sf}
\begin{split}
\frac{d\Sigma_{\mathrm{sf}}}{d\Omega}
&= \frac{8\pi^3}{V} b_{\mathrm{H}}^2
\Big[
|\widetilde{M}_x|^2
+ |\widetilde{M}_y|^2 \cos^4\theta
+ |\widetilde{M}_z|^2 \sin^2\theta \cos^2\theta
\\
&\quad
- (\widetilde{M}_y \widetilde{M}_z^*
+ \widetilde{M}_y^* \widetilde{M}_z)
\sin\theta \cos^3\theta
\Big] ,
\end{split}
\end{equation}
where $V$ is the scattering volume, $b_{\mathrm{H}} = 2.91 \times 10^8 \, \mathrm{A}^{-1}\mathrm{m}^{-1}$ is the magnetic scattering length in the small-angle regime (the atomic magnetic form factor is approximated by $1$, since we are dealing with forward scattering), $\widetilde{\mathbf{M}}(\mathbf{q}) = \{ \widetilde{M}_x, \widetilde{M}_y, \widetilde{M}_z \}$ denotes the Fourier transform of the magnetization vector field $\mathbf{M}(\mathbf{r}) = \{ M_x, M_y, M_z \}$, $\theta$ is the angle between $\mathbf{H}_0 = H_0 \mathbf{e}_z$ and $\mathbf{q}$, so that $\mathbf{q} \simeq q \{ 0, \sin\theta, \cos\theta \}$ in small-angle approximation, and the asterisks ``$*$'' mark the complex conjugate. Note that the polarization-dependent chiral function has been neglected in Eq.~(\ref{eq:sf}), because the experimentally accessible spin-flip sum corresponds to the average of the two spin-flip channels and thereby removes the antisymmetric chiral contribution. Independent of this formal cancellation in the summed channel, the simulated ensembles are also close to balanced in chirality: for each \((f,k)\) case, the fraction with \(\chi=+1\) lies between \(0.44\) and \(0.56\) for \(N=300\) NFs.

The Fourier components $\widetilde{M}_{x,y,z}(\mathbf{q})$ are numerically computed from the results $M_{x,y,z}(\mathbf{r})$ of the micromagnetic (real-space) simulations using the NuMagSANS software package~\cite{adamsjac2026}. The 2D spin-flip SANS cross section [Eq.~(\ref{eq:sf})] depends on $q_y$ and $q_z$, or equivalently on $q=(q_y^2 + q_z^2)^{1/2}$ and $\theta=\operatorname{arctan2}(q_y,\, q_z)$. These direct numerical evaluations provide the reference SANS landscape in Fig.~\ref{fig5}(a). Note that vortex-vortex interparticle dipolar interactions are not included in the present forward model; the analysis is restricted to dilute ensembles of independently simulated NFs.


\subsection{Analytical vortex model}

The interpretation of the micromagnetic textures is guided by the analytical vortex-SANS framework introduced for individual vortex particles in Ref.~\cite{adamsprb2024no2} and by the corresponding angular anisotropy landscape of vortex ensembles developed in Ref.~\cite{adams2026angular}. Both studies provide the reduced language used here: in the leading linear-vortex projection, the spin-flip SANS response of vortex-state particles is organized by the radial vortex profile and by the distribution of vortex-axis orientations. Our micromagnetic analysis therefore does not start from the full 3D magnetization field alone, but first projects each simulated state onto collective variables used in the analytical vortex-SANS theory.

For each simulated remanent state $\mathbf{M}(\mathbf{r};H_0=0)$, the vortex parameter $\nu\ge 0$, the chirality  \(\chi = \pm 1\), the polar angle $\alpha$ and the azimuthal angle $\beta$ were extracted by fitting (detailed in the Supplemental Material~\cite{smelisim2026}) the rotated analytical hyperbolic vortex magnetization vector field:
\begin{align}
\mathbf{M}_{\mathrm{hv}}(\mathbf{r}) &= \mathbf{R}(\alpha,\beta)\mathbf{M}_{\mathrm{hv}}'(\mathbf{R}^T(\alpha,\beta)\mathbf{r}), 
\label{eq:hyperbolic_vortex_main}
\\
\mathbf{M}_{\mathrm{hv}}'(\mathbf{r}')
 &=
 M_{\mathrm{s}}\operatorname{sech}\left(\nu \frac{\rho'}{R}\right) \mathbf{e}_{z'}
 +
\chi M_{\mathrm{s}}\tanh\left(\nu \frac{\rho'}{R}\right) \mathbf{e}_{\phi'} ,\nonumber
\end{align}
where $\mathbf{R}=\mathbf{R}_z(\beta)\mathbf{R}_y(\alpha)$ is the rotation matrix, \(R\) is the particle radius, $\rho' = \sqrt{(x')^2 + (y')^2}$ is the radial cylinder coordinate, $\mathbf{e}_{\phi'}=\{-y'/\rho',+x'/\rho', 0\}$ is the azimuthal unit vector, and $M_{\mathrm{s}}$ is the saturation magnetization. Here, the primed coordinates, e.g., $\mathbf{r}'$, denote the local coordinates, whereas the unprimed coordinates refer to the global coordinates. The parameter \(\nu\) controls the radial vortex profile: a larger \(\nu\) corresponds to a narrower vortex core, a smaller \(\nu\) to a broader vortex core, and \(\nu=0\) to a uniform magnetization state. Because \(N=300\) NFs were simulated for each \((f,k)\) pair, the fitting procedure yielded \(N=300\) corresponding parameter sets \((\nu,\chi,\alpha,\beta)\).

The functional form of the fitted hyperbolic profile is connected to the closed analytical SANS expressions of Refs.~\cite{adamsprb2024no2,adams2026angular} through the corresponding linear-vortex approximation. It is obtained as the small-\(\rho'/R\) limit of Eq.~\eqref{eq:hyperbolic_vortex_main},
\begin{align}
\mathbf{M}_{\mathrm{lv}}(\mathbf{r}) &= \mathbf{R}(\alpha,\beta)\mathbf{M}_{\mathrm{lv}}'(\mathbf{R}^T(\alpha,\beta)\mathbf{r}),
\label{eq:linear_vortex_main}
\\
\mathbf{M}_{\mathrm{lv}}'(\mathbf{r}')
&=
M_{\mathrm{s}}\mathbf{e}_{z'}
+M_{\mathrm{s}}\chi \nu
\frac{ \rho'}{R} \mathbf{e}_{\phi'} .
\nonumber
\end{align}
Within the linear-vortex representation, \(\nu\) fixes the relative weight of the circulating contribution. The purpose of this projection is not to approximate the complete hyperbolic texture pointwise throughout the particle. Rather, it maps the leading vortex symmetry onto a magnetization field that is polynomial in the spatial coordinates, so that its Fourier transform and orientational ensemble average remain analytically closed. At the level of the local small-\(\rho'/R\) expansion, \(\nu/R\) is the first-order radial slope. The independently optimized real-space and reciprocal-space radial parameters used below, however, solve different global projection problems and therefore need not coincide numerically.

Following the notation of Refs.~\cite{adamsprb2024no2,adams2026angular}, with \(u=qR\),
\(F(u)=(\sin u-u\cos u)/u^3\), and \(W=3V_{\mathrm{s}}^2b_{\mathrm{H}}^2 M_{\mathrm{s}}^2/V\), the resulting $(\nu,\chi,\alpha,\beta)$-ensemble-averaged spin-flip SANS cross section is (details in the Supplemental Material~\cite{smelisim2026}):
\begin{align}
\left\langle
\frac{d\Sigma_{\mathrm{sf}}}{d\Omega}
\right\rangle_{\mathrm{lv}}
&=
\frac{W}{8}
\left[F(qR)\right]^2
\mathcal{A}_0(\theta)
\nonumber\\
&\quad
+
\frac{W}{2}\nu_{\mathrm{rms}}^2
\left[F'(qR)\right]^2
\mathcal{A}_1(\theta),
\label{eq:linear_vortex_sf_main}
\end{align}
where the angular profile functions are defined as:
\begin{align}
\mathcal{A}_0(\theta)
&=
12
-
\Lambda
\left(
3\cos^2 2\theta
+
2\cos 2\theta
+
3
\right)
+
4\cos 2\theta,
\nonumber\\
\mathcal{A}_1(\theta)
&=
3
-
\left(
2\Lambda-1
\right)
\cos 2\theta .
\label{eq:linear_vortex_angular_factors_main}
\end{align}
For the ensemble average in Eq.~\eqref{eq:linear_vortex_sf_main}, the variables $(\nu,\chi,\alpha,\beta)$ are assumed to be statistically independent. In addition, we assume a monodisperse particle radius \(R\), equal probabilities for the two circulation senses, and a uniform distribution of $\beta$ on the interval $[0,2\pi)$. The remaining polar-axis distribution is described by the \(\alpha\)~dependent density \(\psi_{\alpha}(\alpha)\), normalized according to \(\int_0^\pi \psi_{\alpha}(\alpha)\sin\alpha\,d\alpha=1\). In contrast to Refs.~\cite{adamsprb2024no2,adams2026angular}, where Eq.~\eqref{eq:linear_vortex_sf_main} was evaluated for a prescribed angular probability density \(\psi(\alpha,\beta)\), we do not impose a specific parametric form on the polar distribution in the main analysis. Instead, its contribution is expressed through the orientational moment:
\begin{align}
\Lambda
=
3\left\langle\cos^2\alpha\right\rangle-1 ,
\label{eq:lambda_main}
\end{align}
where the average is explicitly defined as:
\begin{align}
\left\langle\cos^2\alpha\right\rangle
=
\int_{0}^{\pi} \psi_{\alpha}(\alpha)\cos^2\alpha\,\sin\alpha\,d\alpha .
\end{align}
\begin{figure}[tb!]
\centering
\includegraphics[width=1.0\columnwidth]{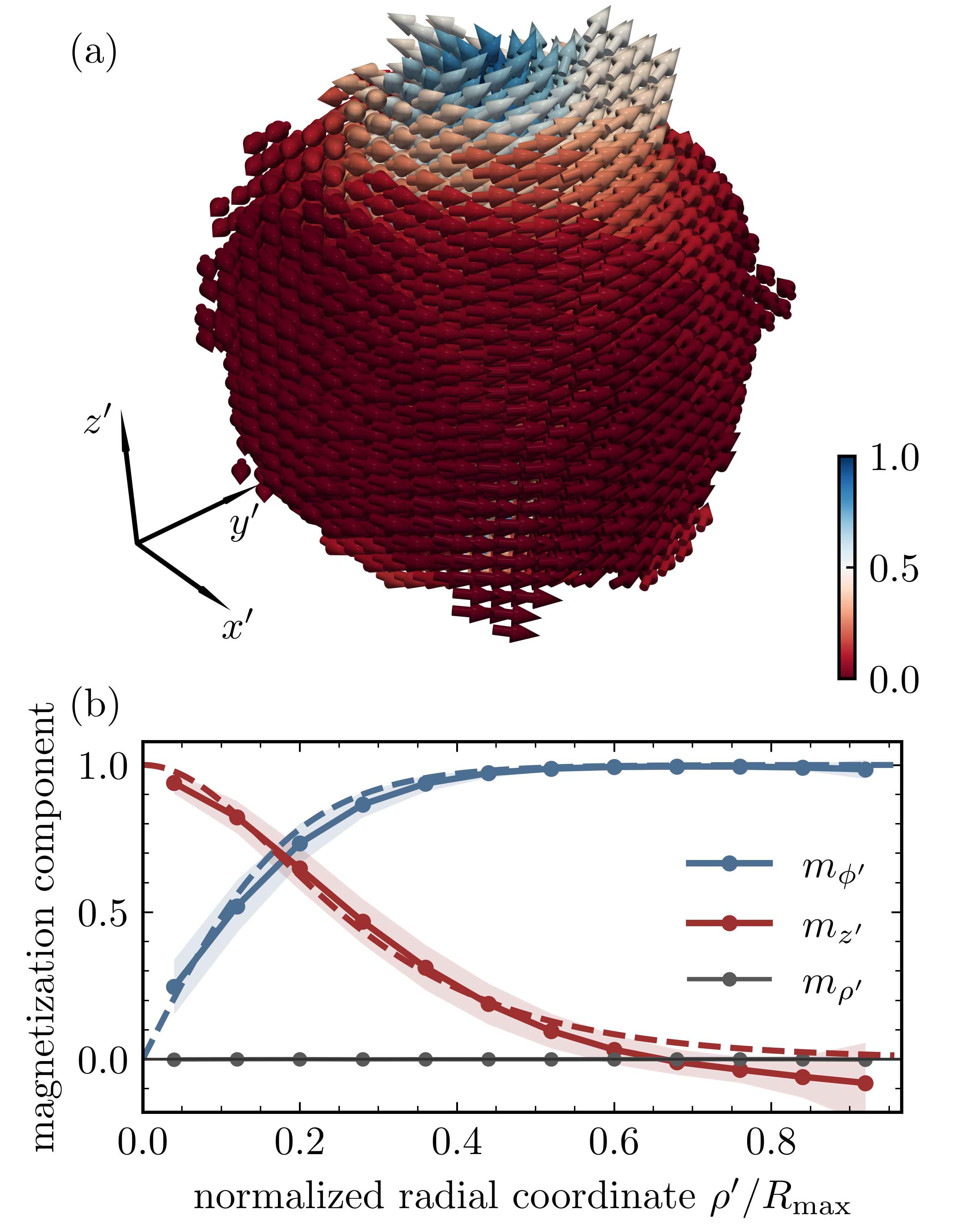}
\caption[Radial vortex profile fit]{Real-space vortex characterization for a
representative remanent NF ensemble with \(f=0.4\), \(k=0.25\), and
\(N=300\). (a)~Three-dimensional magnetization vector field of a
representative NF after transformation into the fitted vortex-axis
coordinate system; color bar encodes the \(m_{z'}\)~component. (b)~Ensemble-averaged radial
profiles in the same coordinate system. Symbols denote the mean binned
micromagnetic profiles, and shaded bands denote three standard deviations
across the ensemble for \(m_{\phi'}\) and \(m_{z'}\). Averaged over the radial
bins, the corresponding \(3\sigma\) half-widths are \(0.033\) and \(0.066\),
respectively. Dashed lines show
the hyperbolic vortex model [Eq.~\eqref{eq:hyperbolic_vortex_main}] evaluated
with \(\nu_{\mathrm{rms},N}=5.26\); the fitted individual values have a standard
deviation of \(0.20\).}
\label{fig3}
\end{figure}
Given that \(0\leq\langle\cos^2\alpha\rangle\leq1\), the orientational moment is bounded by \(-1\leq\Lambda\leq2\). The lower bound corresponds to a purely transverse axis distribution,
\(\psi_{\alpha}(\alpha)=\delta(\alpha-\pi/2)\), yielding \(\Lambda=-1\). The upper bound is reached when all orientational weight is concentrated at \(\alpha=0\) and/or \(\alpha=\pi\), yielding \(\Lambda=2\). By contrast, \(\Lambda=0\) is obtained both for an isotropic distribution, \(\psi_{\alpha}(\alpha)=1/2\), and for a uniform distribution restricted to either the northern or southern hemisphere. Thus, distinct polar-axis distributions can have the same value of \(\Lambda\), because this moment does not uniquely determine \(\psi_{\alpha}(\alpha)\). For the finite micromagnetic ensemble, the corresponding empirical moment is estimated directly from the extracted polar angles as:
\begin{align}
\Lambda_N
&=
\frac{1}{N}
\sum_{j=1}^{N}
\left(3\cos^2\alpha_j-1\right),
\label{eq:lambda_empirical_main}
\end{align}
where \(\alpha_j\) is the fitted polar vortex-axis angle of the \(j\)th NF. The Supplemental Material compares the empirical polar distributions $\psi_{\alpha}(\alpha)$ with uniform-band and linear-segment representations~\cite{smelisim2026}. Their limited accuracy for some parameter combinations motivates the direct use of \(\Lambda\) in the main analysis, without imposing a specific functional form for \(\psi_{\alpha}(\alpha)\).

In the linear-vortex expression, the radial parameter enters the spin-flip SANS cross section quadratically. We therefore summarize the independently fitted real-space values \(\nu_j\) by the empirical root-mean-square value $\nu_{\mathrm{rms},N}$:
\begin{align}
\nu_{\mathrm{rms},N}^{2}
&=
\left\langle\nu^2\right\rangle_N
=
\frac{1}{N}\sum_{j=1}^{N}\nu_j^2 .
\label{eq:nu_rms_main}
\end{align}
This quantity is the real-space radial descriptor corresponding to the quadratic radial coefficient in the reduced SANS model; it is used below as a reference quantity and is not inserted into the independent SANS fits.
For the real-space fit of the \(j\)th NF, \(R\) is set to
the object-specific maximum radial extent \(R_{\max,j}\), as defined in the
Supplemental Material~\cite{smelisim2026}. For the comparison in
Fig.~\ref{fig5}(c), we use the arithmetic ensemble mean
\(\langle R_{\max}\rangle_{N}=N^{-1}\sum_{j=1}^{N}R_{\max,j}
=62.79\,\mathrm{nm}\).

Figure~\ref{fig3} illustrates the real-space reduction for one representative parameter combination. The agreement of \(m_{\phi'}\) and \(m_{z'}\) with Eq.~\eqref{eq:hyperbolic_vortex_main} shows that the fitted hyperbolic model captures the dominant radial structure of the simulated vortex-like textures. Equation~\eqref{eq:linear_vortex_sf_main} provides the analytical basis for the best-fit SANS representations in Fig.~\ref{fig5}(b). The independently extracted real-space descriptors \(\nu_{\mathrm{rms},N}\) and \(\Lambda_N\) provide the reference quantities for the comparisons in Figs.~\ref{fig5}(c,d). The following section examines how these two descriptors depend on the NF parameters \(k\) and \(f\).


\section{Results and Discussion}
\label{results}

\subsection{Collective variables}

We first analyze the radial vortex profile through \(\nu_{\mathrm{rms},N}=\sqrt{\langle\nu^2\rangle_N}\), where \(\nu\) is fitted for each NF from Eq.~\eqref{eq:hyperbolic_vortex_main}. We then summarize the vortex-axis distribution by the empirical moment \(\Lambda_N\) defined in Eq.~\eqref{eq:lambda_empirical_main}. Both quantities are evaluated for each simulated pair of micromagnetic parameters \((k,f)\).

\begin{figure}[t!]
\centering
\includegraphics[width=1.0\columnwidth]{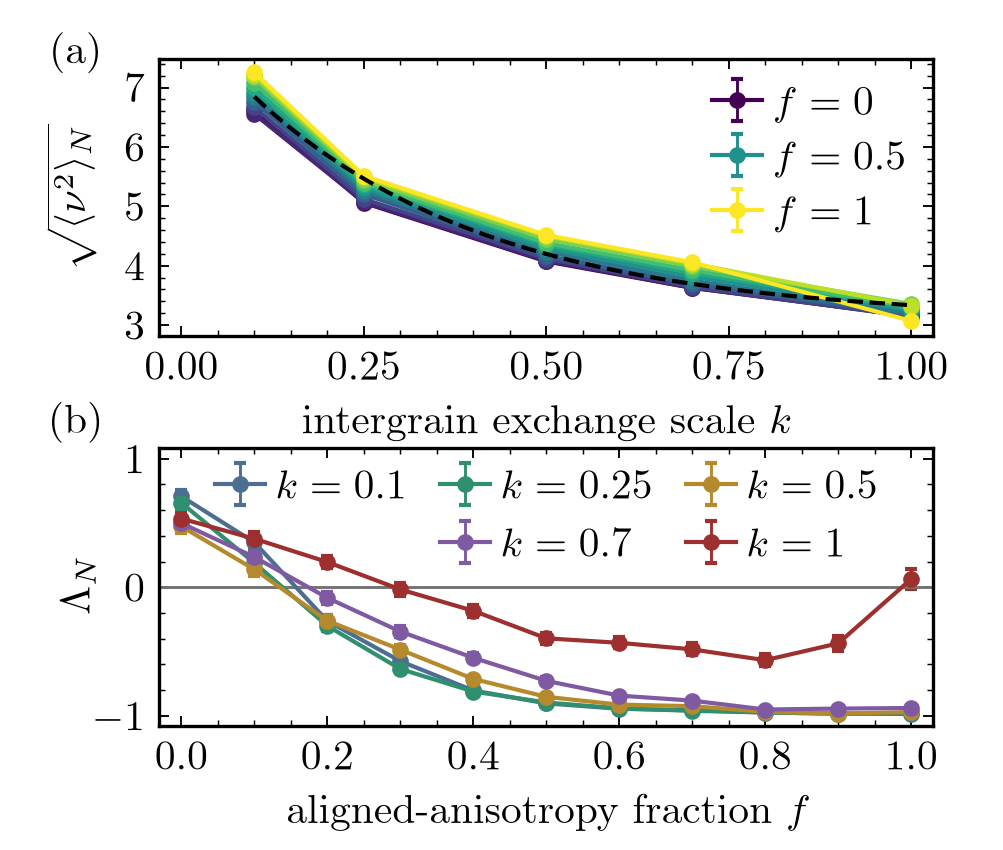}
\caption[Vortex collective variables]{Collective vortex descriptors extracted
directly from the remanent micromagnetic states. (a)~Root-mean-square (RMS) ensemble value
\(\nu_{\mathrm{rms},N}=\sqrt{\langle\nu^2\rangle_N}\) [Eq.~\eqref{eq:nu_rms_main}] of the fitted radial vortex-profile parameter as a function of intergrain exchange scale \(k\). Colors denote different
aligned-anisotropy fractions \(f\), and the dashed line shows a common \(x_1+x_2\exp(-x_3 k)\) fit to all points. (b)~Empirical orientational moment \(\Lambda_N\) [Eq.~\eqref{eq:lambda_empirical_main}], computed directly from the fitted vortex axes for each \((f,k)\). Positive \(\Lambda_N\) denotes axes biased toward the field direction, while negative \(\Lambda_N\) denotes predominantly transverse vortex axes.}
\label{fig4}
\end{figure}

Figure~\ref{fig4} shows that the two extracted descriptors respond differently
to the two micromagnetic control parameters $(k,f)$. The ensemble root-mean-square radial vortex-profile parameter \(\nu_{\mathrm{rms},N}\) decreases with increasing intergrain exchange scale \(k\), which corresponds to an increasing vortex-core broadening, and follows roughly an exponential decay with \(k\). Its dependence on \(f\) is weaker over the sampled parameter range. Qualitatively, reducing \(k\) weakens exchange-driven alignment across the grain boundaries relative to the magnetostatic tendency toward flux closure, consistent with the narrower fitted vortex cores at small \(k\). By contrast, the orientational moment \(\Lambda_N\) varies primarily with \(f\). For \(k=0.1\) to \(0.7\), increasing the fraction $f$ of grains with \(\mathbf{u}_{\mathrm{ani},i}=\mathbf{u}_{\mathrm{ani},0}\) drives \(\Lambda_N\) from positive values, corresponding to axes biased toward the field direction, to values close to \(-1\), corresponding to predominantly transverse vortex axes for $f\gtrsim 0.5$. The \(k=1\) curve is displaced from this collapse, indicating that very strong intergrain exchange partially suppresses the transverse locking of the vortex axes. For the fixed, near-threshold particle size considered here, a qualitative interpretation is that a vortex axis transverse to \(\mathbf{e}_z\) allows part of the circulating magnetization to acquire components along the coherent easy-axis direction, whereas a field-parallel vortex has its circulating component mainly perpendicular to \(\mathbf{e}_z\). The observed vortex-axis reorientation may therefore reflect competition among coherent uniaxial anisotropy, exchange, and dipolar energy in the remanent state, rather than a purely geometrical effect.

The positive-\(\Lambda_N\) region also connects directly to the uniaxial iron-sphere simulations reported in Ref.~\cite{adamsprb2024no2}. These simulations followed the major hysteresis loop from magnetic saturation through remanence. Within the analytical cone model used in Ref.~\cite{adamsprb2024no2}, the high-field limit is described by a vanishing vortex amplitude $\nu$ and \(\alpha_{\mathrm{c}}\rightarrow0^\circ\), whereas the vortex-axis distribution broadens as the field is reduced. At remanence, the vortex-axis ensemble obtained for uniaxial anisotropy was represented by a uniform spherical-cap distribution with \(\alpha_{\mathrm{c}}=54^\circ\). For this distribution, the orientational moment is
\begin{align}
\Lambda &= 
\cos\alpha_{\mathrm{c}} 
+
\cos^2\alpha_{\mathrm{c}}
\simeq 0.93,
\end{align}
corresponding to a pronounced field bias and a vertical two-fold spin-flip SANS anisotropy.

More generally, for a uniform spherical-cap distribution defined by \(0\leq\alpha\leq\alpha_{\mathrm{c}}\), the orientational moment is given by the same expression. Even when \(\alpha_{\mathrm{c}}\) is formally extended over the full interval \(0\leq\alpha_{\mathrm{c}}\leq\pi\), this parametrization is restricted to
\begin{align}
-\frac{1}{4} \leq \Lambda \leq 2,
\end{align}
with its minimum reached at \(\alpha_{\mathrm{c}}=2\pi/3\). The spherical-cap model therefore cannot represent the strongly transverse regime with \(\Lambda_N\simeq-1\) found in the present simulations. The direct use of the empirical orientational moment \(\Lambda_N\) removes this restriction without imposing a specific functional form on the vortex-axis distribution.

Formally, \(\alpha_{\mathrm{c}}\rightarrow0^\circ\) corresponds to \(\Lambda\rightarrow2\). At complete saturation, however, the vortex amplitude $\nu$ vanishes and the vortex axis is no longer an independent physical degree of freedom. Together with the field dependence of the radial vortex-profile parameter demonstrated using the hyperbolic vortex model in Ref.~\cite{adamsprb2026minimal}, this shows that both reduced descriptors can evolve along the magnetic-field cycle. The present parameter grid contains the field-biased remanent regime at small \(f\) and weak intergrain exchange, but additionally extends into the predominantly transverse regime with \(\Lambda_N\simeq-1\).

Thus, in this remanent protocol, \(k\) is mainly associated
with the radial vortex-profile channel, whereas \(f\) is mainly associated
with the vortex-orientation channel. This separation is not an assumed
property of the model, but an outcome of the present micromagnetic ensemble.
It motivates the use of \(\nu_{\mathrm{rms},N}\) and \(\Lambda_N\) as
real-space reference descriptors for comparison with the effective parameters
obtained from the normalized SANS fits.

The Supplemental Material~\cite{smelisim2026} gives the complete \(\psi_{\alpha}(\alpha)\) grids and compares uniform-band and linear-segment representations. These compact distributions are useful for visualizing the axis statistics, but some parameter points are not faithfully represented by such simple densities. The main-text analysis therefore uses \(\Lambda_N\) as the empirical real-space orientational descriptor and compares it directly with \(\Lambda_{\mathrm{fit}}\), while the detailed \(\psi_{\alpha}(\alpha)\) modeling is retained as a diagnostic layer.


\subsection{Two-dimensional spin-flip SANS}
\label{2D}

\begin{figure*}[t!]
\centering
\includegraphics[width=\textwidth,height=0.60\textheight,keepaspectratio]{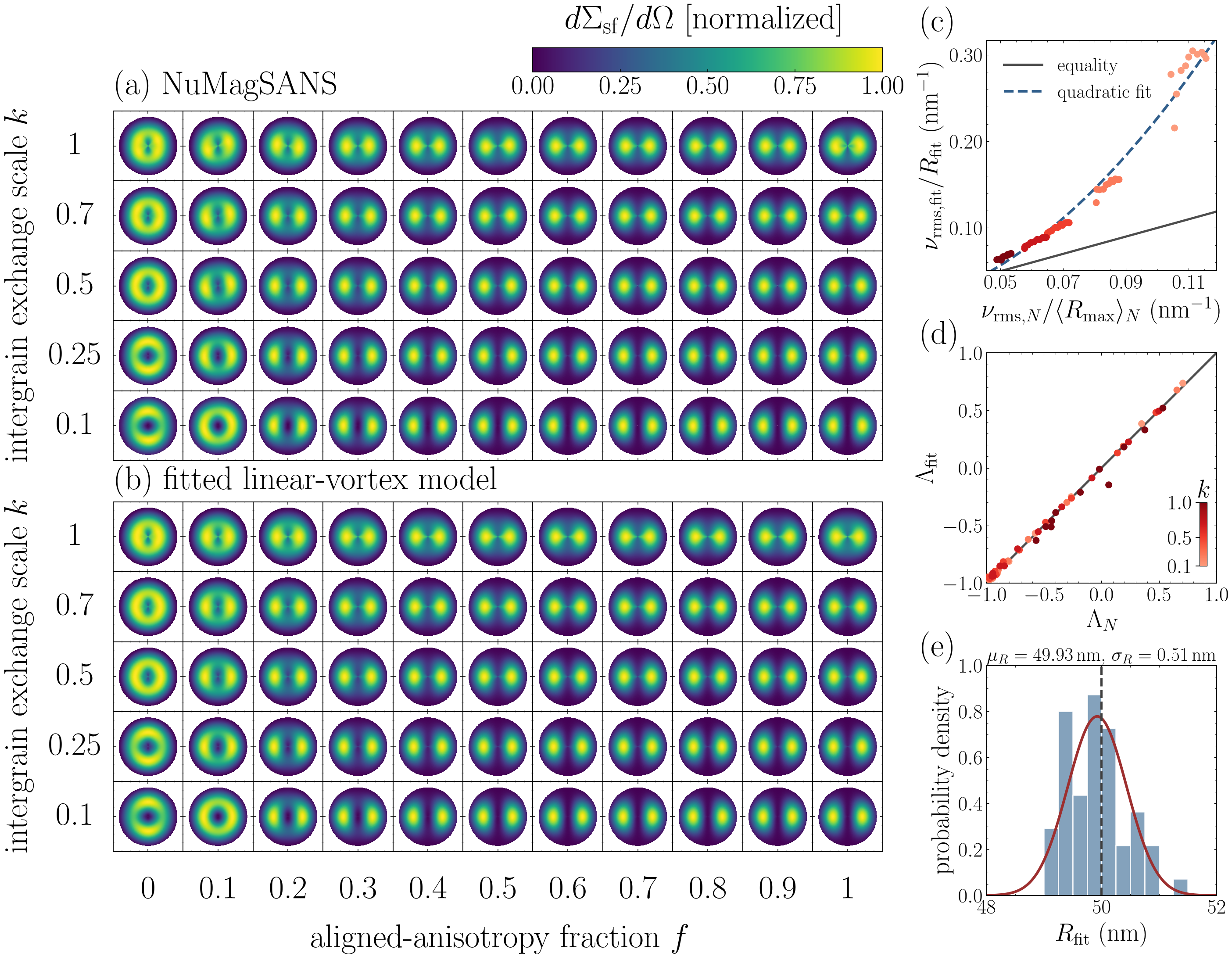}
\caption[Numerical and analytical spin-flip SANS landscapes]{Comparison of the numerical and reduced analytical 2D spin-flip SANS landscapes at remanence. The applied magnetic field $\mathbf{H}_0$ is horizontal in the plane, where the horizontal coordinate is $q_z$ and the vertical one is $q_y$. (a)~NuMagSANS cross sections~\cite{adamsjac2026} computed from the micromagnetic magnetization fields of \(N=300\) independently generated NFs. (b)~Best-fit linear-vortex SANS cross sections obtained independently for each \((f,k)\) using \(\nu_{\mathrm{rms,fit}}\), \(\Lambda_{\mathrm{fit}}\), and \(R_{\mathrm{fit}}\). The horizontal grid axis gives the fraction \(f\) of grains with \(\mathbf{u}_{\mathrm{ani},i}=\mathbf{u}_{\mathrm{ani},0}\), while the vertical grid axis gives the intergrain exchange scale \(k\). The patterns in (a) and (b) are displayed for \(q\leq0.10\,\mathrm{nm}^{-1}\), corresponding to \(qR\leq5\) for the nominal radius of \(R_{\mathrm{nom}}=50\,\mathrm{nm}\), and are normalized separately to compare their shapes rather than absolute intensities. (c)~Comparison of the real-space radial inverse scale \(\nu_{\mathrm{rms},N}/\langle R_{\max}\rangle_{N}\) with the SANS-fit radial inverse scale \(\nu_{\mathrm{rms,fit}}/R_{\mathrm{fit}}\); the solid line denotes equality and the dashed line the global quadratic fit. (d)~Comparison of \(\Lambda_{\mathrm{fit}}\) with the empirical real-space moment \(\Lambda_N\); the line denotes equality and colors indicate \(k\). (e)~Distribution of \(R_{\mathrm{fit}}\) with a Gaussian fit.}
\label{fig5}
\end{figure*}

Figure~\ref{fig5} compares the numerical spin-flip SANS cross sections
calculated from the micromagnetic magnetization fields using
NuMagSANS~\cite{adamsjac2026} [Fig.~\ref{fig5}(a)] with the corresponding
best fits of the linear-vortex SANS model
[Fig.~\ref{fig5}(b)]. For each \((f,k)\), the parameters
\(\nu_{\mathrm{rms,fit}}\), \(\Lambda_{\mathrm{fit}}\), and
\(R_{\mathrm{fit}}\) were obtained by minimizing the pixelwise error between
the individually normalized numerical and analytical cross sections.
Consequently, the comparison tests the ability of the linear-vortex model
to represent the radial and angular pattern shape, but neither its absolute
intensity nor a parameter-free forward prediction from the real-space
descriptors. The mean normalized RMSE is \(0.0124\).

The radial scales extracted from the
SANS and real-space analyses show a strong global nonlinear relation
[Fig.~\ref{fig5}(c)], whereas
\(\Lambda_{\mathrm{fit}}\) closely follows the real-space moment
\(\Lambda_N\) [Fig.~\ref{fig5}(d)]. This orientational agreement is nontrivial because the numerical cross sections originate from the full irregular micromagnetic textures rather than from the reduced linear-vortex model. Nevertheless, \(\Lambda_{\mathrm{fit}}\) follows \(\Lambda_N\) with a Pearson correlation coefficient of \(0.998\) and an RMSE of \(0.0368\), despite the moderate deviations from the factorization assumption documented in the Supplemental Material~\cite{smelisim2026}. The nonlinear relation in Fig.~\ref{fig5}(c) represents a global empirical
trend over all sampled \((f,k)\) combinations. It is not assumed to hold
separately along fixed-\(f\) or fixed-\(k\) subsets and therefore does not
define a unique pointwise conversion between the real-space and SANS-fit
radial descriptors. The fitted radii are narrowly
distributed around the nominal NF radius, with
\(\langle R_{\mathrm{fit}}\rangle=49.93\,\mathrm{nm}\) and
\(\sigma_R=0.51\,\mathrm{nm}\) [Fig.~\ref{fig5}(e)].
The underlying real-space 
vortex-axis distributions and additional comparisons with explicit 
models for \(\psi_{\alpha}(\alpha)\) are documented in the 
Supplemental Material~\cite{smelisim2026}. 

The radial evolution of the numerical SANS cross sections can be summarized by
azimuthal averaging,
\begin{align}
I(q)
=
\frac{1}{2\pi}
\int_0^{2\pi}
\left\langle\frac{d\Sigma_{\mathrm{sf}}}{d\Omega}\right\rangle(q,\theta)\,d\theta .
\end{align}
For each \((f,k)\), the peak position \(q_{\max}\), peak intensity
\(I_{\max}\), and relative central intensity \(I(0)/I_{\max}\) are extracted
from this radial average. Table~\ref{tab:radial_sans_characteristics} compares
their averages over the sampled values of \(f\) with the corresponding
real-space descriptor \(\langle\nu_{\mathrm{rms},N}\rangle_f\). With
increasing \(k\), \(\langle\nu_{\mathrm{rms},N}\rangle_f\) decreases from
\(6.915\) to \(3.220\), while \(\langle q_{\max}R_{\mathrm{nom}}\rangle_f\) decreases from
\(2.462\) to \(1.924\) and \(\langle I(0)/I_{\max}\rangle_f\) increases from
\(0.059\) to \(0.724\). As shown in Fig.~\ref{fig4}, the variation over \(f\) is small compared with this
systematic dependence on \(k\). For weak intergrain exchange,
\(\langle q_{\max}R_{\mathrm{nom}}\rangle_f=2.462\) lies close to the maximum of
\(\lvert F'(qR)\rvert\) at \(qR\simeq2.5\) in the linear-vortex model.
Increasing \(k\) shifts the numerical maximum toward smaller \(q\) and raises
the relative central intensity, consistent with the vortex-core broadening
indicated by the decreasing radial descriptor in Fig.~\ref{fig4}(a).
\begin{table}[h!]
\caption{Comparison of the real-space radial vortex descriptor with radial
characteristics of the azimuthally averaged numerical spin-flip SANS cross
sections. The brackets denote averages over the sampled values of \(f\), and
\(I_{\max}=\max_q I(q)\).}
\label{tab:radial_sans_characteristics}
\begin{ruledtabular}
\begin{tabular}{c c c c}
\(k\)
&
\(\langle\nu_{\mathrm{rms},N}\rangle_f\)
&
\(\langle q_{\max}R_{\mathrm{nom}}\rangle_f\)
&
\(\left\langle I(0)/I_{\max}\right\rangle_f\)
\\
\hline
0.10 & 6.915 & 2.462 & 0.059 \\
0.25 & 5.293 & 2.377 & 0.212 \\
0.50 & 4.289 & 2.238 & 0.427 \\
0.70 & 3.822 & 2.123 & 0.561 \\
1.00 & 3.220 & 1.924 & 0.724 \\
\end{tabular}
\end{ruledtabular}
\end{table}
The evolution of the angular anisotropy can be interpreted through the
angular factors \(\mathcal A_0(\theta)\) and
\(\mathcal A_1(\theta)\) in
Eq.~\eqref{eq:linear_vortex_angular_factors_main}. We therefore
consider three characteristic parameter regions separately.

In the following limiting-case discussion, the values of
\(\nu_{\mathrm{rms},N}\) and \(\Lambda_N\) from Fig.~\ref{fig4} are used to
locate the different \((f,k)\) regions and to characterize their real-space
vortex profiles qualitatively. The generic parameter
\(\nu_{\mathrm{rms}}\) appearing in the analytical expressions below denotes
the radial coefficient of the linear-vortex SANS model and is not identified
point by point with the real-space estimator
\(\nu_{\mathrm{rms},N}\).
\paragraph{Parameter region \(f\gtrsim0.5\),
\(0.1\lesssim k\lesssim0.7\).}---As shown in Fig.~\ref{fig4}, for \(f\gtrsim0.5\) and \(0.1\lesssim k \lesssim0.7\), the orientational moment approaches \(\Lambda_N\simeq-1\), indicating a predominantly transverse vortex-axis distribution. Within this parameter region, the radial vortex descriptor lies approximately in the interval \(4\lesssim\nu_{\mathrm{rms},N}\lesssim7\). For the limiting case \(\Lambda=-1\), the analytical SANS cross section~\eqref{eq:linear_vortex_sf_main} has the finite Fourier representation:
\begin{align}
\left\langle
\frac{d\Sigma_{\mathrm{sf}}}{d\Omega}
\right\rangle_{\Lambda=-1}
&=
I_0(q)
+
I_2(q)\cos 2\theta
+
I_4(q)\cos 4\theta ,
\label{eq:sf_fourier_transverse}
\end{align}
with the \(q\)~dependent intensities (for $\Lambda=-1$):
\begin{align}
I_0(q)
&=
\frac{3W}{2}
\left[
\frac{11}{8}\left[F(qR)\right]^2
+
\nu_{\mathrm{rms}}^2
\left[F'(qR)\right]^2
\right],
\nonumber\\
I_2(q)
&=
\frac{3W}{2}
\left[
\frac{1}{2}\left[F(qR)\right]^2
+
\nu_{\mathrm{rms}}^2
\left[F'(qR)\right]^2
\right],
\nonumber\\
I_4(q)
&=
\frac{3W}{16}\left[F(qR)\right]^2.
\label{eq:sf_fourier_coefficients_transverse}
\end{align}
At \(\theta=\{0,\;\pi\}\), the second-order harmonic yields \(\cos2\theta=1\), whereas at \(\theta=\{\pi/2, \;3\pi/2\}\) it is \(\cos2\theta=-1\). The fourth-order harmonic has the same value, \(\cos4\theta=1\), in both directions. Since \(I_2(q)>0\), the spin-flip cross section is therefore larger at \(\theta=\{0,\;\pi\}\), producing the field-parallel pair of lobes observed in Fig.~\ref{fig5}.

The real-space interval
\(4\lesssim\nu_{\mathrm{rms},N}\lesssim7\) identifies this part of the
\((f,k)\) map as a pronounced vortex-profile regime. It is used here as a
qualitative reference for interpreting the angular limiting case. Within the
fitted linear-vortex representation, the relative radial weight is instead
set by \(\nu_{\mathrm{rms,fit}}\) and remains \(q\) dependent through
\([F'(qR)]^2\), which vanishes at \(q=0\) and reaches its maximum near
\(qR\simeq2.5\)~\cite{adamsprb2024no2}.

\paragraph{Parameter region \(f=0\), \(0.1\lesssim k\lesssim1.0\).}--- In the small-\(f\) region, represented here by \(f=0\), Fig.~\ref{fig5} shows a transition from a vertical two-fold anisotropy for \(k=0.1\) and \(0.25\) to a nearly ringlike or weakly horizontal two-fold anisotropy for \(k\geq0.5\). As shown in Fig.~\ref{fig4}, the corresponding orientational moment lies approximately within \(0.4\lesssim\Lambda_N\lesssim0.8\), while the radial descriptor covers the range \(3\lesssim\nu_{\mathrm{rms},N}\lesssim7\). The real-space range
\(3\lesssim\nu_{\mathrm{rms},N}\lesssim7\) is used here to characterize the
corresponding radial-profile regime. The angular evolution of the fitted
linear-vortex representation can then be examined through
\(\mathcal A_1(\theta)\), particularly near the maximum of
\(\lvert F'(qR)\rvert\) at \(qR\simeq2.5\)~\cite{adamsprb2024no2}. For \(\Lambda=1\),
\begin{align}
\mathcal A_1(\theta,\Lambda=1)
&=
3-\cos 2\theta ,
\end{align}
which has its maxima at \(\theta=\pi/2\) and \(3\pi/2\) and therefore exhibits a vertical two-fold anisotropy. For \(\Lambda=0\),
\begin{align}
\mathcal A_1(\theta,\Lambda=0)
&=
3+\cos 2\theta ,
\end{align}
which has its maxima at \(\theta=0\) and \(\pi\) and therefore exhibits a horizontal two-fold anisotropy. At the intermediate value \(\Lambda=1/2\),
\begin{align}
\mathcal A_1(\theta,\Lambda=1/2)
&=
3,
\end{align}
so that \(\mathcal A_1(\theta)\) is isotropic at this value. Its two-fold anisotropy is vertical for \(\Lambda>1/2\) and horizontal for \(\Lambda<1/2\). This interpretation of the transition at \(\Lambda=1/2\) should be understood within the high-\(\nu\) limit of the reduced vortex model, in which the \(F'\)~dependent term governs the angular anisotropy over the relevant finite-\(q\) range. This corresponds to the limiting-case analysis used to organize the four characteristic regions of the analytical vortex-SANS landscape in Ref.~\cite{adams2026angular}. For finite values of \(\nu_{\mathrm{rms,fit}}\), however, the complete fitted linear-vortex cross section is not generally isotropic at \(\Lambda=1/2\), because the \(F\)~dependent angular factor \(\mathcal A_0(\theta)\) retains second- and fourth-order angular harmonics.

\paragraph{Parameter region \(0.1\lesssim f\lesssim 0.4\), \(0.1\lesssim k\lesssim1.0\).}--- In the intermediate-\(f\) region, the orientational moment decreases from moderately positive to negative values and covers approximately \(-0.8\lesssim\Lambda_N\lesssim0.4\), as shown in Fig.~\ref{fig4}. The radial descriptor remains primarily controlled by \(k\) and lies within \(3\lesssim\nu_{\mathrm{rms},N}\lesssim7\). Although \(\Lambda_N\) passes through zero within this region, it remains below \(1/2\). Consequently, \(\mathcal A_1(\theta)\) retains a horizontal two-fold anisotropy throughout this part of the parameter grid.

The case \(\Lambda=0\) illustrates why passing through zero does not produce an isotropic spin-flip SANS pattern. In this case, the angular factors reduce
to
\begin{align}
\mathcal A_0(\theta,\Lambda=0)
&=
4\left(3+\cos2\theta\right),
\nonumber\\
\mathcal A_1(\theta,\Lambda=0)
&=
3+\cos2\theta ,
\end{align}
and Eq.~\eqref{eq:linear_vortex_sf_main} becomes:
\begin{align}
\left\langle
\frac{d\Sigma_{\mathrm{sf}}}{d\Omega}
\right\rangle_{\Lambda=0}
&=
I_0(q)+I_2(q)\cos2\theta ,
\end{align}
where, for \(\Lambda=0\), the \(q\)~dependent intensities are now
\begin{align}
    I_0(q) &= \frac{3W}{2}\left[\left[F(qR)\right]^2
    +
    \nu_{\mathrm{rms}}^2\left[F'(qR)\right]^2\right] ,
    \\
    I_2(q) &= \frac{W}{2}\left[\left[F(qR)\right]^2
    +
    \nu_{\mathrm{rms}}^2\left[F'(qR)\right]^2\right] .
\end{align}
An isotropic vortex-axis distribution yields \(\Lambda=0\), although this value is not unique to that distribution. The corresponding cross section nevertheless retains a horizontal two-fold anisotropy with maxima at \(\theta=0\) and \(\pi\). As shown in Fig.~\ref{fig4}, \(\Lambda_N\) decreases further toward \(-1\) as \(f\) increases, while the principal lobe direction remains field-parallel and the angular contrast and fourth-order harmonic change. This explains why the intermediate-\(f\) patterns in Fig.~\ref{fig5} can appear similar to the \(f\gtrsim0.5\) patterns despite their different values of \(\Lambda_N\).


\section{Conclusion}
\label{conclusion}

Using polycrystalline iron oxide multicore nanoparticles as an example, we have shown how micromagnetic simulations, polarized SANS, and analytical vortex descriptors can be combined in a forward model that maps intraparticle disorder onto collective vortex variables. The nanoflower structures were generated as 3D Voronoi region maps, and the magnetic disorder was varied through two explicit model parameters: the intergrain exchange scale \(k\) and the fraction \(f\) of grains whose anisotropy axes are aligned with a common reference direction.

The central outcome is that the vortexlike remanent textures are organized by two collective channels. Within the present forward model, the dominant variation of \(k\) is associated with the radial vortex-profile parameter \(\nu_{\mathrm{rms}}\), whereas the dominant variation of \(f\) is associated with the vortex-axis moment \(\Lambda=3\langle\cos^2\alpha\rangle-1\). The full polar distribution \(\psi_{\alpha}(\alpha)\) remains useful for visualizing the fitted axes, yet its leading projection into the linear-vortex SANS cross section is \(\Lambda\). Consequently, the micromagnetic parameter space can be interpreted as a mapping of \((k,f)\rightarrow(\nu_{\mathrm{rms}},\Lambda)\), which positions the simulated nanoflowers in characteristic regions of the analytical vortex-SANS landscape. Here, this micromagnetic mapping refers specifically to the real-space estimators \((\nu_{\mathrm{rms},N},\Lambda_N)\); the independently fitted SANS parameters quantify how the same textures are represented within the linear-vortex SANS cross section.

This separation is also visible directly in reciprocal space. For weak intergrain exchange, the azimuthally averaged numerical SANS cross section exhibits a pronounced finite-\(q\) maximum with \(q_{\max}R_{\mathrm{nom}}\simeq2.5\). With increasing \(k\) and decreasing \(\nu_{\mathrm{rms},N}\), this maximum shifts toward smaller \(q_{\max}R_{\mathrm{nom}}\), while the relative central intensity increases, consistent with a broadening vortex core. In parallel, the variation of \(\Lambda_N\) organizes the transition between vertical two-fold, nearly ringlike, and horizontal two-fold anisotropies. Independent fits of the linear-vortex model reproduce the normalized numerical SANS morphologies with a mean RMSE of \(0.0124\). The orientational moments obtained from the SANS and real-space analyses agree closely, whereas the corresponding radial scales exhibit a global nonlinear relation.

The present \((k,f)\) map should therefore be understood as a fixed-size remanent section through a broader vortex-descriptor space. Previous size-dependent nanoflower simulations distinguish core-dominated and flux-closure-dominated reversal regimes~\cite{eliss2026}, whereas the minimal spherical-vortex model demonstrates that the radial profile parameter evolves substantially along the magnetic-field cycle~\cite{adamsprb2026minimal}. Particle size, magnetic field, and field protocol may therefore provide additional routes through the same \((\nu_{\mathrm{rms}},\Lambda)\) descriptor space. The relationships identified here are consequently conditional on the selected particle size and remanent protocol rather than constituting a size-independent classification.

The comparison also clarifies the associated inverse problem. For textures independently identified as vortexlike in real space, the fitted orientational moment \(\Lambda_{\mathrm{fit}}\) closely follows the empirical moment \(\Lambda_N\). This near identity between independently determined quantities provides direct numerical support for \(\Lambda\) as the leading orientational information channel retained by the reduced SANS description. By contrast, the SANS-fitted and real-space radial scales exhibit a global nonlinear relation, so the former should not be interpreted as a direct reconstruction of the latter. Inferring the full vortex-axis distribution or microscopic parameters such as \(k\) and \(f\) requires additional model information, since similar reduced descriptors may arise from particle size, magnetic field, field protocol, or other structural parameters.

A second level of non-uniqueness concerns the model class itself. Because the real-space ensemble is fully known in the present study, we can trace how its collective information propagates through object-wise vortex fitting, ensemble averaging, Fourier projection, and normalized SANS fitting. This provides a controlled test of the linear-vortex model as a reduced representation, but does not establish that this model can be uniquely identified from SANS data alone. Denoting the normalized fitting residual of a model class schematically by \(\varepsilon\), the present result \(\varepsilon_{\mathrm{lv}}\ll1\) does not imply \(\varepsilon_{\mathrm{alt}}\gg1\) for every alternative model class. It demonstrates representability and consistency of the linear-vortex reduction for a known vortex ensemble, but not unique model selection from a normalized SANS cross section.

We propose as a next step to go beyond the leading linear-vortex projection used here. Although the present treatment shows that the leading orientational moment accounts for a large part of the variation in the normalized 2D SANS cross section, the remaining deviations may encode higher-order radial terms and higher spherical moments of the vortex-axis distribution. Such an extension would be conceptually consistent with the second-order power-series analysis introduced previously for the azimuthally averaged magnetic SANS cross section~\cite{adamsprb2024}. However, that framework also illustrates the cost of this route: even for the 1D azimuthal average, the second-order description involves a set of seven intensity coefficients rather than a single additional parameter. A vortex-specific second-order model would therefore have to balance the improved description of the \(q\)~dependence against a substantially less constrained inverse problem. In this sense, the leading descriptors used here provide
the controlled baseline on top of which higher-order corrections can be added.

The model-comparison statements made here refer to individually normalized SANS cross sections. Absolute intensity calibration could provide additional information, but would also introduce further experimental nuisance parameters, including concentration, magnetic volume, saturation magnetization, particle-size dispersion, and instrumental resolution.


\section*{Data availability}
The micromagnetic input data, generated nanoflower microstructures, simulation outputs, and analysis scripts associated with this work are available from the Zenodo dataset in Ref.~\cite{adams2026nanoflowerdataset}.

\section*{Acknowledgments}

M.P.A.\ acknowledges support by the European Union
(Grant Agreement No.\ 101135546, MaMMoS).
E.M.J.\ acknowledges funding from the European Union's Horizon 2020 research and innovation program under the Marie Sk{\l}odowska-Curie Actions grant agreement 101081455---YIA and from the Institute for Advanced Studies (IAS) of the University of Luxembourg. J.L.\ is supported by the Ghent University special research fund (bof/baf/1y/2024/01/005 and bof/baf/1y/2025/01/017). The computational resources and services used in this work were provided by the HPC facilities of the University of Luxembourg and the Luxembourg National Supercomputer MeluXina.
The views and opinions expressed are, however, those of
the authors only and do not necessarily reflect those of the
European Union or the European Health and Digital Executive Agency (HADEA).
Neither the European Union nor the
granting authority can be held responsible for them.


\bibliography{references}

\end{document}